\documentclass[biblatex, english]{lni}
\usepackage{booktabs}
\usepackage[]{blindtext}
\graphicspath{{figures/}}
\usepackage{multirow}
\usepackage{subcaption}
\usepackage{graphicx}

\usepackage{tikz}
\usepackage{graphicx}
\usetikzlibrary{arrows.meta, positioning, fit, calc, shapes.geometric}

\usepackage{xargs}                      
\usepackage[pdftex,dvipsnames]{xcolor}  
\usepackage[colorinlistoftodos,prependcaption,textsize=tiny]{todonotes}
\newcommandx{\unsure}[2][1=]{\todo[linecolor=red,backgroundcolor=red!25,bordercolor=red,#1]{#2}}
\newcommandx{\change}[2][1=]{\todo[linecolor=blue,backgroundcolor=blue!25,bordercolor=blue,#1]{#2}}
\newcommandx{\info}[2][1=]{\todo[linecolor=OliveGreen,backgroundcolor=OliveGreen!25,bordercolor=OliveGreen,#1]{#2}}
\newcommandx{\improvement}[2][1=]{\todo[linecolor=Plum,backgroundcolor=Plum!25,bordercolor=Plum,#1]{#2}}
\newcommandx{\thiswillnotshow}[2][1=]{\todo[disable,#1]{#2}}

\begin{document}
\title[Ein Kurztitel]{Adaptable multimodal Key Information Extraction from Screenshots: Combining Vision and Language Information}
\author[1]{Arjun Nepal}{}{}
\author[1]{Tobias Bocklet}{}{}

\affil[1]{Technische Hochschule Nürnberg Georg Simon Ohm\\Center for Artificial Intelligence\\ Keßlerplatz 12\\90489 Nuremberg\\Germany}
\maketitle

\begin{abstract}
Structured information extraction from visually rich documents is increasingly important for automating document-centric workflows, where relevant information is often embedded in screenshots from mobile applications, wearable devices, and machine interfaces. This paper investigates multimodal key information extraction from heterogeneous sports and fitness screenshots by comparing three paradigms: layout-aware OCR-based extraction, document-oriented vision encoder–decoder models, and parameter-efficiently adapted multimodal large language models (MLLMs). Experimental results show that incorporating visual semantics significantly improves extraction robustness over purely rule OCR-based approaches, while document-oriented generative models enable competitive OCR-free extraction. The best performance is achieved by a Low-Rank Adaptation (LoRA)-adapted MLLM, reaching $88.1\%$ All slots recall and outperforming both OCR-based and document-oriented approaches by approximately $8-10\%$. Furthermore, increased LoRA adaptation capacity yields additional gains, confirming the effectiveness of parameter-efficient multimodal adaptation. These findings demonstrate that adapted multimodal foundation models provide a robust and transferable solution for information extraction from heterogeneous visual documents and heterogeneous screenshot-based document understanding tasks applications.
\end{abstract}


\section{Introduction}
The extraction of structured information from visually rich documents is a fundamental problem in document understanding, as documents contain not only textual content but also layout, style, and visual elements that influence interpretation~\cite{LayoutLM}. Key Information Extraction (KIE) addresses this challenge by identifying relevant information fields from document images and converting them into structured representations. Prior work formulates KIE as a downstream task of OCR, but also highlights that textual features alone are insufficient, as document understanding requires incorporating both visual semantics and layout structure~\cite{PICK}.

A central difficulty in KIE lies in modeling relationships between entities in a two-dimensional document space. Unlike sequence-based information extraction, the association between a key and its value depends on spatial arrangement, reading order, and surrounding visual context. This requirement for joint modeling of textual, spatial, and visual information has been emphasized in works such as~\cite{BROS} and further supported by layout-aware approaches including~\cite{LayoutLMV2}, which integrate multimodal signals for document understanding.

Early OCR-centered systems typically separate text recognition from downstream extraction, leading to error propagation and reduced robustness under varying document layouts and languages~\cite{Donut}. To address these limitations, recent approaches move towards multimodal and generative formulations that directly operate on document images. These advances have been facilitated by large-scale benchmarks such as DocVQA~\cite{DocVQA}, which provide diverse document images and question-answer pairs for evaluating document understanding capabilities. In particular, models such as~\cite{GenKIE} reformulate KIE as a sequence-to-sequence problem, reducing reliance on token-level annotations while incorporating visual and layout information. 

Despite these advances, existing methods still face challenges in robustly handling heterogeneous and visually complex documents. This is especially evident in born-digital data such as mobile user interfaces and device screens, where relevant information may appear not only as text but also as icons, widgets, numerical indicators, and spatially organized visual elements. Prior work has shown that such interfaces require holistic multimodal understanding, as their semantics are distributed across text, structure, and visual components~\cite{Pix2Struct, Screen2Words}.

The automatic extraction of structured information from visually rich documents is increasingly important for small and medium-sized enterprises (SMEs), where information is often distributed across reports, forms, screenshots, application interfaces, and scanned documents. Automating this process can reduce manual effort and support downstream analytics and business workflows. However, real-world document collections are highly diverse, containing varying layouts, structures, document types, OCR quality, and domain-specific semantics. As a result, KIE systems must generalize across heterogeneous documents and extraction schemas while requiring minimal task-specific adaptation~\cite{Kleister,RealKIE,BuDDIE}.

This paper addresses the extraction of structured key-value pairs (KVPs) from sports and fitness activity screenshots collected from mobile applications, wearable devices, and fitness-machine interfaces. The objective is to automatically extract five activity-related attributes: activity type, date, duration, distance, and calories. Although these target attributes remain consistent across sources, they are expressed through diverse visual cues, including textual labels, icons, graphical widgets, and prominently displayed numerical values. Additional challenges arise from multilingual interfaces, incomplete metadata, relative temporal expressions, and the presence of multiple semantically related measurements within a single screenshot. Accurate extraction therefore requires not only identifying relevant information but also correctly interpreting and associating visual and semantic cues across varied interface designs.

To address these challenges, three complementary paradigms for multimodal KIE are investigated. The first follows a layout-aware extraction strategy that combines OCR-derived text, spatial reasoning, and icon detection to exploit both textual and visual cues. The second explores document-oriented encoder–decoder architectures, including established document understanding models and custom vision encoder–language decoder combinations, which generate structured outputs directly without handcrafted extraction rules from document images. The third leverages MLLMs in both zero-shot and parameter-efficient fine-tuning settings, enabling instruction-guided extraction through pretrained vision–language reasoning capabilities. Together, these paradigms facilitate a systematic comparison between traditional extraction pipelines, task-specific generative document models, and adaptable multimodal foundation models for structured information extraction from heterogeneous visually rich documents.

The main contributions of this work are summarized as follows: 
\begin{itemize}
    \item A domain-specific case study of multimodal KIE on heterogeneous sports and fitness activity screenshots.

    \item A systematic comparison of OCR-based systems, document-oriented generative models, and multimodal large language models.

    \item An assessment of parameter-efficient adaptation of multimodal foundation models through LoRA, including an analysis of the trade-off between adaptation capacity and extraction performance.
\end{itemize}

The remainder of this paper is organized as follows. Section~\ref{sec:rel_work} reviews related work on key information extraction, document understanding, and multimodal vision-language models. Section~\ref{sec:dataset} describes the dataset and evaluation protocol. Section~\ref{sec:methodology} presents the investigated extraction approaches. Section~\ref{sec:results} reports experimental results, comparative analyses, and the LoRA ablation study. Finally, Section~\ref{sec:conclusion} concludes the paper and outlines directions for future work. 

\section{Related Work} 
\label{sec:rel_work}
\textbf{OCR-based Systems, Heuristic Extraction, and Visual Semantic Cues :}
OCR-based KIE systems typically represent documents as sets of text tokens with bounding boxes, where structured information is extracted using spatially grounded rules over these tokens. Constraint-based approaches such as ~\cite{BluePrint} formulate extraction as applying spatial and semantic constraints (e.g., proximity, alignment, ordering) over OCR outputs to associate keys and values, which directly supports strategies such as searching for candidate values along horizontal or vertical axes around detected keys. Hybrid systems further combine learned models with deterministic rules, where neural predictions are refined using domain-specific constraints such as formatting, positional consistency, or keyword matching~\cite{PandLRuleKIE, ReceiptIEHybrid}. In parallel, document layout analysis methods introduce visual semantics by detecting structural elements such as tables, headers, and text blocks using object detection models~\cite{PubLayNet, DocBank}, enabling extraction systems to map visual regions or detected objects to semantic roles when textual cues are incomplete. These approaches form the basis of practical KIE systems but rely on predefined rules or mappings, limiting their robustness across diverse document layouts.

\textbf{Multimodal Relationship Extraction:}
Learning-based KIE methods replace heuristic matching with explicit modeling of relationships between document entities. Early approaches represent documents as graphs, where textual entities are connected through spatial and semantic relationships and contextual information is propagated using graph neural networks~\cite{Katti2018, PICK}. Subsequent work extends this idea through adaptive graph refinement, multimodal relation alignment, and spatial-aware graph structures to capture both local and global dependencies between document elements~\cite{GraphRevisedIE, Zheng2021, DocTr}. Other approaches formulate key--value extraction as an explicit entity-linking problem, where semantic relationships between keys and values are directly predicted through attention-based decoding~\cite{KVPFormer}. Collectively, these methods demonstrate that modeling structural and semantic relationships is essential for robust KIE beyond token-level classification.

\textbf{Layout-aware Document Models:}
Layout-aware document models extend OCR-based representations by incorporating spatial and visual information directly into document embeddings. The LayoutLM family integrates textual content with layout coordinates and visual features to jointly model text, layout, and image information within Transformer architectures~\cite{LayoutLM, LayoutLMV2, LayoutLMv3}. Subsequent approaches explore alternative representations of document structure through multimodal feature fusion, structure-aware pretraining, graph-based structural modeling, and multilingual layout representations ~\cite{DocFormer, StructuralLM, FormNet, LiLT, LayoutXLM}. By preserving geometric and structural dependencies between document elements, these models substantially improve document understanding and KIE performance compared with text-only representations, although they generally remain dependent on OCR-derived textual tokens.

\textbf{Generative Document Models and Vision-Language Models:} Generative document models reformulate document understanding as sequence generation using encoder--decoder architectures. ~\cite{Donut} employs a Swin Transformer encoder to process document images and a BART-style autoregressive decoder to generate structured outputs directly from visual inputs, eliminating the need for OCR. ~\cite{Pix2Struct} adopts an image encoder as normal Vision Transformer (ViT) and text decoder trained to generate structured representations such as simplified HTML from screenshots, enabling parsing of visually rich documents and interfaces. Pix2Struct also addresses limitations of fixed-resolution vision encoders by preserving variable document layouts through adaptive patch extraction, enabling more faithful modeling of visually structured content such as forms, screenshots, and HTML-rendered interfaces that are common in KIE tasks. ~\cite{GenKIE} formulates KIE as a multimodal sequence-to-sequence task by combining visual, layout, and textual encoders with a Transformer decoder, where constrained prompting is used to enforce structured output generation. ~\cite{Udop} extends this paradigm by unifying text, layout, and image modalities within an encoder--decoder framework for multiple document understanding tasks with visual grounding. 

Vision-language models (VLMs) further generalize this idea by integrating image encoders with large language models (LLMs). ~\cite{DocLLM} incorporates layout-aware embeddings into a generative language model and modifies attention mechanisms to integrate spatial information during decoding. Instruction-tuned models such as~\cite{LayoutLLM} adapt large multimodal models for document understanding by framing extraction as question answering through instruction-tuning. Models such as Qwen2.5-VL and Qwen3-VL explicitly support document parsing, visual grounding, and structured extraction over long multimodal contexts, making them suitable candidates for flexible KIE across heterogeneous visual documents~\cite{qwen25vl, qwen3vl}. Full fine-tuning of large multimodal models is costly because all parameters must be updated and stored for each task. SmolVLM introduces architectural and tokenization optimizations that enable strong image and document understanding capabilities at a model scale of approximately 2.2B parameters and significantly reduced memory requirements compared with larger VLMs~\cite{smolvlm}. LoRA addresses this by learning low-rank adapters on top of frozen pretrained weights, enabling parameter-efficient adaptation with comparable performance for domain-specific finetuning~\cite{LoRA, LoRA_v}.

Taken together, prior work shows three complementary directions for KIE: explicit OCR-based extraction, document-specific generative modeling, and instruction-guided multimodal foundation models. This motivates the methodology of this paper, which evaluates these three paradigms under a unified domain-specific KIE/KVP extraction setting.

\section{Dataset} 
\label{sec:dataset}
The dataset consists of screenshots of sports and fitness activities collected from multiple mobile applications, wearable-devices and fitness-machine interfaces, with evaluation performed on a test set of 998 annotated images. The training (4,195) and validation (1,399) sets were synthetically generated from approximately 1102 real screenshots by replacing slot values while preserving the original layout and key–value structure, whereas the 998 sample test set contains real-world screenshots. The test set covers 14 different tracker categories.
\begin{figure}[h!]
    \centering
    \includegraphics[width=\textwidth, height=7cm]{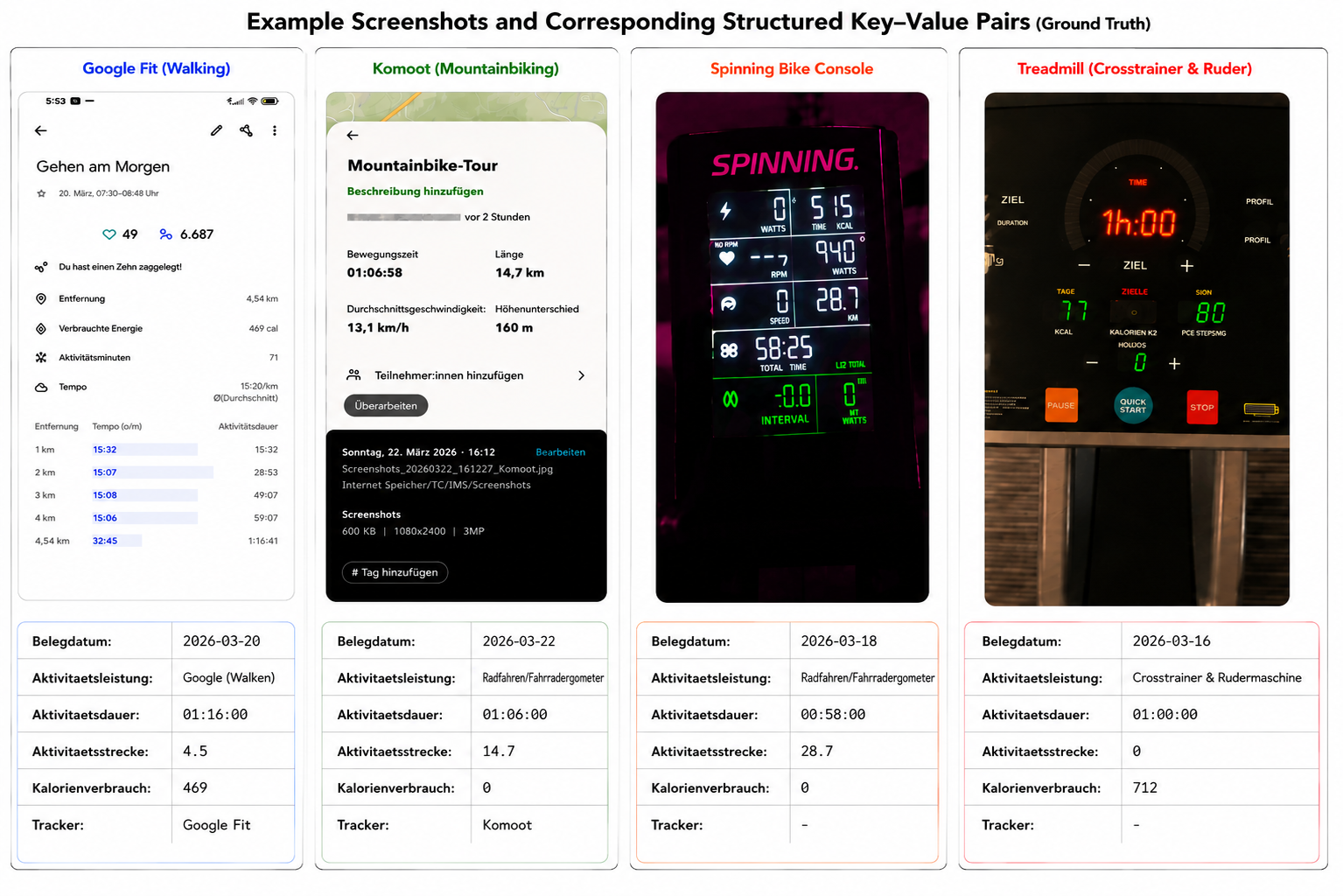}
    \caption{Example screenshots with corresponding ground truth KVP.}
    \label{fig:dataset_examples}
\end{figure}
\figurename~\ref{fig:dataset_examples} presents representative screenshots and their ground-truth KVP. The examples highlight challenges in dataset creation, where annotations often require normalization, correction, and semantic interpretation rather than direct extraction. For example, a displayed duration of \textit{1:16:41} is normalized to \textit{01:16:00} although the screenshots also contains \textit{Aktivitaetsminuten 71}, while a relative timestamp such as \textit{vor 2 Stunden} must be resolved to the corresponding activity date. Furthermore, visually diverse activity descriptions are mapped to a unified label set, and the correct slot values must be selected from multiple semantically related measurements displayed within the same interface.

\begin{figure}[h!]
\centering
\includegraphics[width=\linewidth,height=7cm]{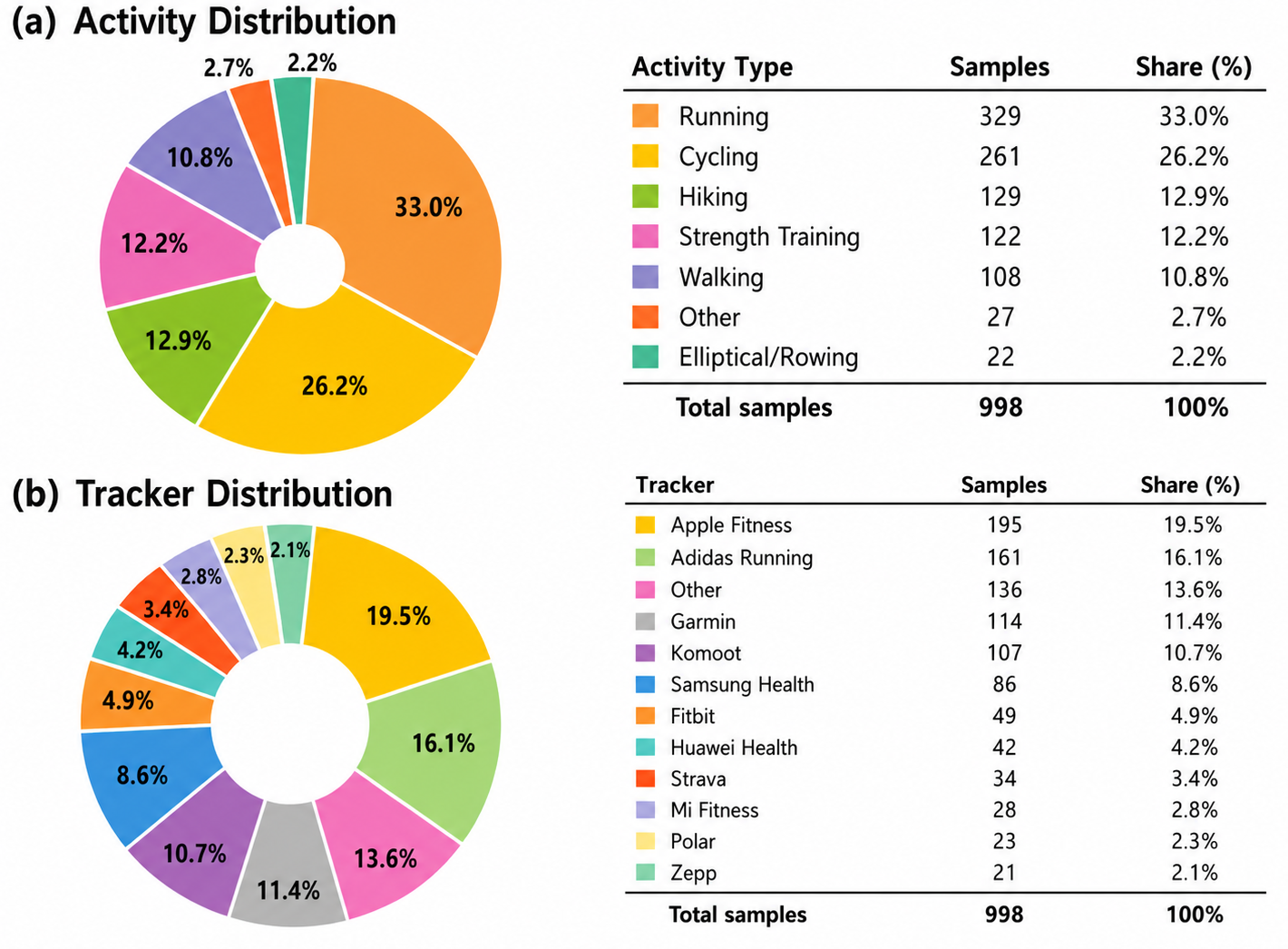}
\caption{Test-set statistics ($n=998$): (a) activity distribution and (b) tracker distribution, categories below 2\% are grouped as $Other$.}
\label{fig:dataset}
\end{figure}

The extraction task comprises five slots: Activity, Date, Duration, Distance, and Calories, with numerical and temporal values extracted verbatim and activity type requiring semantic interpretation from visual and textual cues. Activity is available for all screenshots, duration and date for 99.9\%, distance for 79.0\%, and calories for 67.9\%. All five ground-truth key--value pairs are present in 468 screenshots (46.9\%), which form the candidate set for the $All$ recall metric. \figurename~\ref{fig:dataset} shows the activity and tracker distributions. The dataset exhibits real-world variability arising from heterogeneous tracker layouts and multilingual interfaces. 
\section{Proposed Method} 
\label{sec:methodology}
\begin{figure}[h!]
    \centering
    \includegraphics[width=\linewidth, height=\textwidth, keepaspectratio]{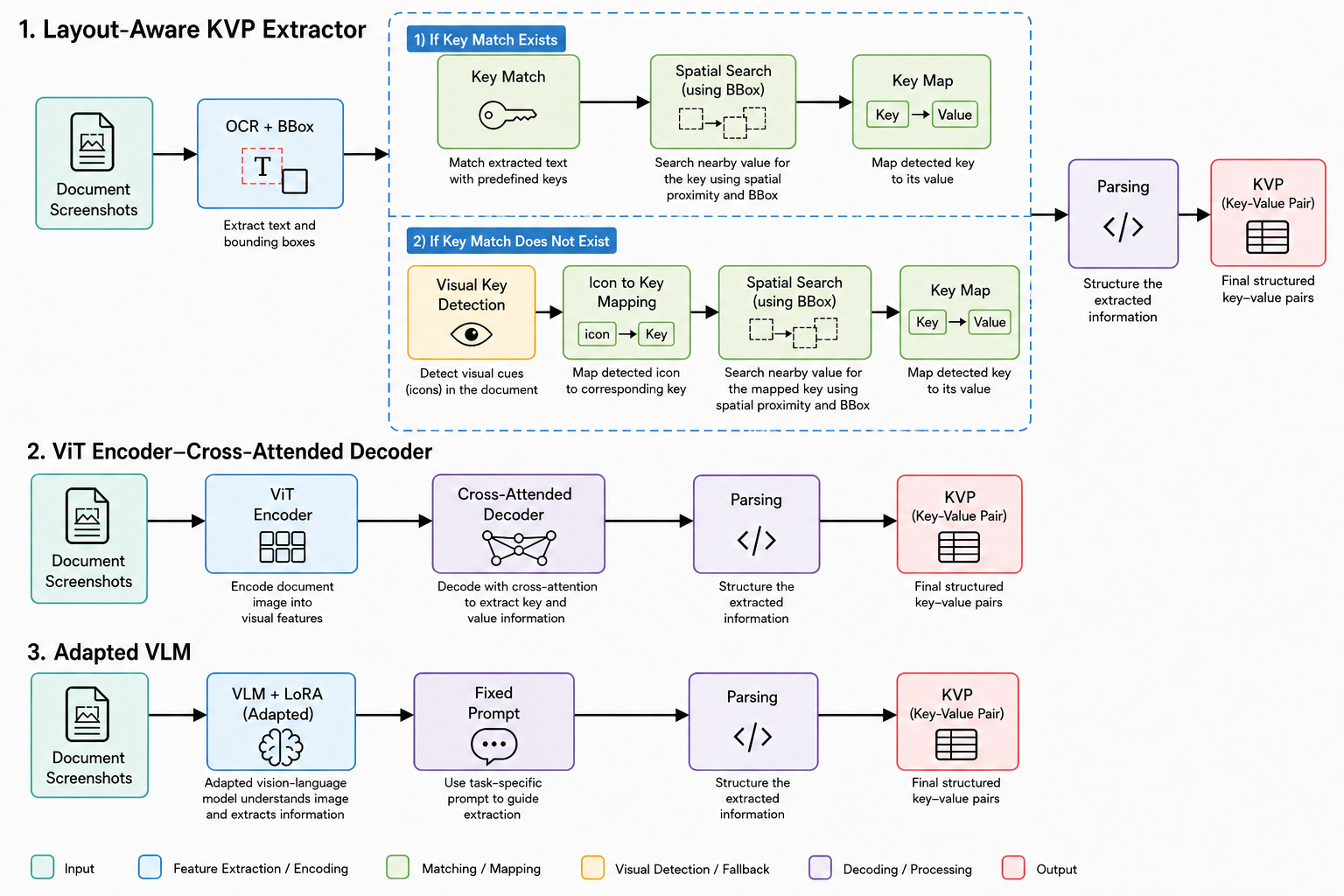}
    \caption{Schematic representation of three document KIE methods.}
    \label{fig:method_schematic}
\end{figure}

Based on detailed analysis of related work, the proposed framework addresses KIE from document screenshots by modeling the task as transforming visual document inputs into structured KVPs. As illustrated in \figurename~\ref{fig:method_schematic}, the overall system is implemented through three methodological paradigms, each reflecting a different level of abstraction in multimodal document understanding. The first paradigm follows a traditional OCR-driven system augmented with spatial reasoning and visual semantics, where text tokens and their geometric layouts are explicitly processed. The second paradigm adopts a learning-based encoder–decoder architecture, where document images are encoded into visual representations and decoded into structured outputs through cross-attention, enabling implicit modeling of key–value relationships. The third paradigm leverages adapted VLM, where extraction is formulated as an fixed instruction-guided generation task, allowing direct mapping from images to structured outputs without explicit intermediate representations. Together, these approaches provide a progressive transition from rule-based extraction to fully generative multimodal reasoning, enabling robustness across diverse document layouts and levels of structural complexity. 
\subsection{Layout-Aware KVP Extractor}
In this method, document images are first processed using OCR to extract textual tokens along with their corresponding bounding box coordinates, forming a spatially grounded representation of the document. The OCR component is based on the PaddleOCR framework and the PP-OCR family of practical text-detection and recognition systems~\cite{paddleocr}. If explicit key matches are found within the extracted text, predefined key vocabularies are used to identify candidate keys, and corresponding values are retrieved through spatial search based on bounding box proximity and alignment. In cases where textual keys are absent, a visual fallback mechanism is employed using a Darknet-based YOLOv4 object detection model trained on document screenshots having sports related icons for activity, date time, duration, distance and calories~\cite{YOLOv4}. YOLOv4 was selected because its open-source Darknet implementation. Detected icons are mapped to their corresponding semantic keys through an icon-to-key mapping scheme, after which the same spatial search strategy is applied to locate associated values. This hybrid design integrates OCR-based structural parsing with learned visual cue detection, enabling the system to handle both text-explicit and visually implied key scenarios before producing the final structured KVP output.

\subsection{Vision Transformer Encoder-Decoder}
Vision Encoder–Decoder architectures generate structured text by conditioning an autoregressive decoder on visual representations through cross-attention. Depending on the encoder, these representations may consist of dense multimodal tokens (Qwen3-VL~\cite{qwen3vl}), hierarchical spatial features (Swin~\cite{swin}), or compact semantic embeddings (SigLIP2~\cite{SigLIP2}), enabling flexible combinations of vision encoders and language decoders for document understanding.

A custom VisionEncoderDecoder model was implemented using the Qwen3-VL-2B vision encoder and GPT-2 decoder~\cite{gpt2}. Since Qwen3-VL produces long visual token sequences $([B,12920,1024])$ where B denotes batch size, a learned query-based resampler compressed the encoder output into a compact latent representation of size $([B,32,1024])$ before cross-attention, reducing the computational cost of autoregressive decoding while preserving salient visual information required for KVP extraction.

Document-oriented baselines included Donut and Pix2Struct using DocVQA-finetuned checkpoints. Donut combines a Swin encoder with a BART decoder~\cite{BART}, while Pix2Struct employs a ViT encoder~\cite{Vit} and T5-style decoder~\cite{T5} with aspect-ratio-preserving image resizing and 2D positional embeddings to better retain document structure. DocVQA-adapted checkpoints were selected to improve document-grounded decoding and multilingual document understanding. A second custom architecture combined SigLIP2-base-patch16-256 with GPT-2. GPT-2 was converted into an encoder-decoder model by enabling cross-attention layers within each transformer block, allowing autoregressive generation conditioned on encoder-derived visual features. Pretrained GPT-2 parameters were retained, while the newly introduced cross-attention layers were initialized and optimized during fine-tuning.

\subsection{Adapted MLLMs \& Zero shot Evaluation}
Both SmolVLM and Qwen3-VL were fine-tuned under the same instruction-tuning framework using identical prompts, dataset splits, training strategy, and LoRA hyperparameters (($r=8$), ($\alpha=8$)), enabling a controlled comparison for the KVP extraction task. In both models, LoRA adapters were applied to the main transformer attention and feed-forward projection layers; however, Qwen3-VL additionally supports dedicated multimodal cross-attention and visual projection modules, allowing adapters to also be inserted into the vision-language interaction components of the architecture. SmolVLM does not expose these explicit cross-attention projection layers in the same modular form, and therefore LoRA adaptation was limited to its shared transformer backbone. A dropout of (0.1) was used for SmolVLM, while Qwen3-VL employed a smaller dropout of (0.01) to maintain stable adaptation across the larger set of multimodal projection layers. Additionally the \textit{Mistral-24B}~\cite{mistral} model was evaluated in a pure zero-shot setting without any adapter-based fine-tuning. The model was prompted directly using the instruction from 
in order to assess the effectiveness of large-scale general-purpose vision-language reasoning without task-specific adaptation. Furthermore canonical activity classes were provided
as additional context during prompting. 

\section{Results \& Discussion} 
\label{sec:results}
To evaluate the effectiveness of the approaches presented for KVP extraction from sports activity screenshots, experiments were conducted using the respective training and validation splits, while all reported results correspond to the held-out test set summarized in \tablename~\ref{table:kvp_results}. The evaluation compares three categories of methods: OCR-based retrieval approaches, Vision Encoder–Decoder architectures, and parameter-efficiently fine-tuned MLLMs, analyzing their ability to extract activity, date, duration, distance, and calorie information. The $All$ metric denotes complete-record recall and is calculated only for screenshots whose ground truth contains all five key–value pairs. It represents the proportion of these complete-ground-truth screenshots for which all five corresponding values are extracted correctly. Screenshots missing one or more ground-truth fields are excluded from this calculation.

\begin{table}[!hb]
\small
\setlength{\tabcolsep}{6pt}
\begin{tabular}{l@{\hspace{10pt}}r@{\hspace{10pt}}r@{\hspace{10pt}}r@{\hspace{10pt}}r@{\hspace{10pt}}r@{\hspace{10pt}}r}
\toprule
\textbf{Method} &
\textbf{Activity} &
\textbf{Date}&
\textbf{Duration} &
\textbf{Distance} &
\textbf{Calories} &
\textbf{All} \\
\midrule
Rule-KVP        & 72.85 & 82.67 & 88.87 & 88.03 & 87.84 & 68.10 \\
Icon-KVP        & 87.17 & 83.77 & 89.55 & 88.06 & 90.24 & 80.13 \\
\midrule
Donut(Swin-BART) & 85.87 & 86.17 & 90.15 & 91.87 & 90.53 & 80.78 \\ 
Pix2Struct(ViT-T5) & 85.87 & 84.97 & 90.45 & 93.52 & 91.12 & 80.78 \\
ViT-CAD (S2-G2) & 86.67 & 67.6 & 90.55 & 91.36 & 77.96 & 52.92 \\
ViT-CAD (Qw-G2) &90.37 & 22.34 & 54.77 & 47.4 & 10.65 & 1.08 \\
\midrule
Mistral-24B      & 79.50 & 81.74 & 73.64 & 91.28 & 88.05 & 49.07\\
LoRA-VLM (S)    &90.08 & 86.37 & 91.26 & 92.76 & 88.76 & 79.70\\
LoRA-VLM (QW)   & 91.88 & 88.88 & 91.96 & 94.28 & 92.75 & 88.12 \\
\bottomrule 
\end{tabular}
\caption{Recall evaluation results in \% for KVP extraction methods. ViT-CAD configurations use S2 = SigLIP2, Qw = Qwen3-VL-vis encoders, G2 = GPT-2, and BA = BART. LoRA-VLM configurations use S = SmolVLM-2.2B and QW = Qwen3-VL-2B.}
\label{table:kvp_results}
\end{table}
Rule-KVP and Icon-KVP demonstrate that OCR-based spatial retrieval is highly effective for semi-structured activity screenshots when stable textual anchors are available. Rule-KVP achieved strong recall on localized numeric fields such as duration, distance, and calories, while Icon-KVP further improved all slots recall by $12.03\%$ and activity extraction by $14.32\%$ by incorporating visual icon detection and semantic key mapping. However, both methods remain heavily dependent on predefined keywords, regex rules, and OCR quality, making them sensitive to language changes, unseen interface variations, and ambiguous unit-based matches that may introduce false positives in visually dense screenshots.

To reduce dependency on predefined textual keys and rule-based spatial heuristics, end-to-end Vision Encoder–Decoder architectures were explored. Donut achieved $90.15\%$, $91.87\%$, and $90.53\%$ recall for duration, distance, and calories respectively, while Pix2Struct achieved $90.45\%$, $93.52\%$, and $91.12\%$. Both models therefore exceeded $90\%$ recall on most structured numeric fields, indicating effective document-grounded decoding from DocVQA-finetuned checkpoints. However, activity recognition remained comparatively weaker at $85.87\%$, particularly for fitness-device screenshots where activities are represented through icons or layout cues rather than explicit text. Preliminary experiments using generic pretrained Donut and Pix2Struct checkpoints produced substantially weaker and unstable structured outputs compared to their DocVQA-finetuned variants, suggesting that downstream document understanding adaptation is critical for reliable evidence localization and document-conditioned generation in KVP extraction tasks. In contrast, replacing the jointly trained Donut decoder with independently pretrained language decoders such as BART and GPT-2 resulted in unstable repetitive outputs, highlighting the importance of learned encoder-decoder alignment for document-conditioned generation. 

Although DocVQA-pretrained architectures produced balanced extraction quality, their weaker activity recognition motivated experiments with stronger multimodal visual encoders and custom cross-attended decoder configurations.
The custom Vision Encoder-Decoder combinations showed highly contrasting behavior despite sharing the same GPT-2 decoder. SigLIP2-GPT2 underperformed for all slots recall of $52.92\%$ and $67.60\%$ for date , while maintaining strong extraction performance for duration and distance at $90.55\%$ and $91.36\%$ respectively. This illustrates that it provides sufficiently strong visual representations for localized KVP retrieval but struggles with extracting exact date from the screenshots. The comparatively lower activity recall of $86.67\%$ suggests that strong global visual representations alone are insufficient for robust activity discrimination. Since many trackers share nearly identical layouts across activities, accurate classification often depends on localized textual cues or small visual content rather than global visual appearance, which may not be fully preserved in the extracted visual representations. In contrast, the Qwen3-VL–GPT2 configuration achieved only $1.08\%$ all slots recall despite strong activity recognition of $90.37\%$. Inspection of the predictions showed that the model often identified the correct document region but generated near-correct dates, durations, or calorie values instead of faithfully reproducing the observed content, suggesting limitations in document-grounded value transcription rather than visual understanding alone.

Among the MLLM-based approaches, the LoRA-adapted Qwen3-VL 2B achieved the best overall performance at $88.12\%$, substantially outperforming both the SmolVLM-based variant at $79.70\%$ and the zero-shot Mistral-24B baseline at $49.07\%$, despite having nearly $12 \times$ fewer parameters than Mistral. The improvement was consistent across all extraction fields, indicating that the model learned a robust alignment between visual elements and structured output representations. Compared to SmolVLM, Qwen3-VL particularly benefited from its stronger multimodal pretraining and vision-language fusion mechanisms, which improved the localization and transcription of document content. In contrast, the considerably lower performance of the zero-shot Mistral baseline highlights that general multimodal reasoning alone is insufficient for reliable schema-constrained KVP extraction without task-specific adaptation. Although the adapted MLLMs achieved the strongest overall extraction performance, they remain computationally expensive during inference due to autoregressive token generation and large multimodal transformer architectures, where inference cost scales with generated sequence length despite parameter-efficient LoRA training.

\subsection{Impact of LoRA Parameterization}
Since the LoRA-adapted Qwen3-VL-2B achieved the strongest overall performance among all evaluated KVP extraction approaches \tablename~\ref{table:kvp_results}, an ablation study was conducted to assess the effect of LoRA parameterization on extraction accuracy. LoRA adapters were applied to the self-attention projections (\texttt{q\_proj}, \texttt{k\_proj}, \texttt{v\_proj}, \texttt{o\_proj}), which control token interactions, the MLP projections (\texttt{up\_proj}, \texttt{down\_proj}), which transform and store semantic representations, and the cross-attention projections responsible for integrating visual and textual information in the multimodal architecture. The LoRA configuration $r_{\alpha}$ was varied while keeping the target modules fixed.

As shown in \tablename~\ref{tab:qwen3vl_lora_ablation}, increasing adapter capacity consistently improves extraction performance. Overall recall rises from $88.34\%$ for configuration $8_{16}$ to $90.06\%$ for configuration $32_{64}$, with the largest model achieving the best results for most fields, including $95.04\%$ recall for Distance extraction. Nevertheless, the gains diminish relative to the increase in trainable parameters. Expanding the adapter from $6.88M$ to $13.76M$ parameters results only a marginal change in overall recall, whereas the largest configuration with $27.53M$ parameters provides a more substantial improvement. These findings indicate that larger LoRA adapters improve extraction quality, albeit with decreasing returns in parameter efficiency.

\begin{table}[!h]
\small
\setlength{\tabcolsep}{6pt}
\begin{tabular}{lrrrrrrrr}
\toprule
\textbf{$r_{\alpha}$} &
\textbf{Params} &
\textbf{Activity} &
\textbf{Date} &
\textbf{Duration} &
\textbf{Distance} &
\textbf{Calories} &
\textbf{All} \\
\midrule
$8_{16}$  & 6.88M  & 91.88 & 88.88 & 91.96 & 94.28 & 92.75 & 88.34 \\
$16_{32}$ & 13.76M & 91.98 & 88.68 & 92.26 & 95.04 & 93.05 & 88.17 \\
$32_{64}$ & 27.53M & 92.39 & 89.38 & 92.56 & 95.04 & 93.20 & \textbf{90.06} \\
\bottomrule
\end{tabular}
\caption{Ablation study of LoRA configurations for Qwen3-VL-2B. Target modules comprise attention projections ($q,k,v,o$), MLP projections (up/down), and cross-attention projections. Params denotes the number of trainable LoRA parameters.}
\label{tab:qwen3vl_lora_ablation}
\end{table}

\section{Conclusion \& Outlook} 
\label{sec:conclusion}
This paper investigated multimodal KIE from heterogeneous sports and fitness activity screenshots using three complementary paradigms: OCR-based extraction with visual cues, document-oriented encoder–decoder architectures, and MLLMs. The result demonstrated that robust extraction requires understanding of multimodal information such as text, visual, layout and spatial information. While OCR-based approaches provide strong performance on explicitly represented textual fields, the incorporation of visual semantics substantially improves the extraction of visually encoded attributes such as activity type. However, icon-dependent solutions remain sensitive to interface modifications and evolving visual designs, requiring additional maintenance and adaptation when application layouts change.

Among the evaluated approaches, encoder–decoder architectures proved capable of performing OCR-free extraction directly from document images, whereas parameter-efficiently adapted multimodal foundation models achieved the most consistent performance across all extraction categories. In particular, activity recognition and semantically ambiguous fields such as calorie and date with relative temporal information benefited from the broader visual-language reasoning capabilities of adapted MLLMs. Furthermore, the LoRA ablation study showed that increasing adaptation capacity improves extraction quality, although the gains diminish relative to the increase in trainable parameters, highlighting an important trade-off between model complexity and efficiency.

Future work will focus on improving the interpretability and robustness of multimodal extraction systems. One promising direction is the integration of visual grounding mechanisms that not only generate structured key–value pairs but also localize the document regions supporting each prediction and provide confidence estimates. Such capabilities could be enabled through lightweight decoder architectures with auxiliary prediction heads for extraction, localization, and uncertainty estimation. In addition, incorporating document layout information and structure-aware reasoning may improve performance on ambiguous or incomplete inputs. Finally, continual adaptation strategies and evaluations across a broader range of document domains will be investigated to improve robustness against interface changes, data drift, and cross-domain deployment scenarios.

\renewcommand{\refname}{Bibliography}
\clearpage
\printbibliography






\end{document}